# Electronic Excitation Response of DNA to High-energy Proton Radiation in Water

Christopher Shepard[1], Dillon C. Yost[3], Yosuke Kanai[1,2*]

1. Department of Chemistry, University of North Carolina at Chapel Hill; Chapel Hill, NC, 27514, USA.

2. Department of Physics and Astronomy, University of North Carolina at Chapel Hill; Chapel Hill, NC, 27514, USA

3. Department of Materials Science and Engineering, Massachusetts Institute of Technology; Cambridge, MA, 02139, USA.

*Corresponding author. **Email:** ykanai@unc.edu

**Abstract:**
The lack of molecular-level understanding for the electronic excitation response of DNA to charged particle radiation, such as high-energy protons, remains a fundamental scientific bottleneck in advancing proton and other ion beam cancer therapies. In particular, the dependence of different types of DNA damage on high-energy protons represents a significant knowledge void. Here we employ first-principles real-time time-dependent density functional theory simulation, using a massively-parallel supercomputer, to unravel the quantum-mechanical details of the energy transfer from high-energy protons to DNA in water. The calculations reveal that protons deposit significantly more energy onto the DNA sugar-phosphate side chains than onto the nucleobases, and greater energy transfer is expected onto the DNA side chains than onto water. As a result of this electronic stopping process, highly energetic holes are generated on the DNA side chains as a source of oxidative damage.

**Introduction**

Understanding the radiation-induced response of DNA is pivotal for human health. The electronic excitation induced in DNA by high-energy protons is of great importance to understanding how DNA damage occurs in extreme conditions such as those experienced by astronauts. For instance, as much as 90% of galactic cosmic radiation (GCR) is high-energy protons, and human exposure to GCR is a great concern for space missions, as limited data exists on the bodily effects [2]. The electronic excitation response of DNA to high-energy protons is also the foundation of modern proton beam cancer therapy. Over the past 30 years, proton beam therapy has emerged as a promising alternative to conventional X-rays in radiation oncology [3]. Having a spatially-localized energy deposition profile, with the so-called Bragg peak being its maximum, the ion beam can more precisely target tumor cells, while minimally affecting surrounding healthy cells [4, 5]. In proton beam therapy, the energy deposition profile needs to be developed for individual patients, and the velocity-dependent energy transfer rate from irradiating protons in water plays a central role [6, 7]. This quantity, often called linear energy transfer or electronic stopping power, is given per unit distance traveled by irradiating protons. The initial kinetic energy of the protons is on the order of a few hundred MeV. As the protons slow down by transferring their momentum, the stopping power increases greatly near the Bragg peak velocity. In addition to having the ideal energy transfer behavior, many studies indicate that proton beams yield complex clustering lesions with strand breaks, including double-strand breaks, as the direct effect on DNA [8]. These strand lesions, particularly with other lesions nearby, are much more likely to lead to cell death [9]. However, how the proton beam induces DNA lesions is not understood at the molecular level [10], and details of the energy transfer mechanism from irradiating protons to DNA are needed to help fill this crucial knowledge gap [11]. The ultrafast nature of the excitations and the need for a particle accelerator, like a cyclotron, to generate high-energy protons makes experimental investigation difficult [12]. On the theory side, perturbation theories based on the dielectric function are widely used, and the current state-of-the-art approach builds on developing an accurate energy-loss function for dry DNA and liquid water as the target [13-16]. Modern quantum-mechanical simulation offers an alternative approach for investigating such electronic stopping phenomena on the molecular level [17, 18]. Non-equilibrium simulations of electron dynamics have significantly benefitted from recent advances in massively-parallel computers with peta- and exa- floating point operations per second performance [19], and unraveling the quantum-mechanical details at the molecular level has impacted various research areas [20, 21]. In particular, with the development of *time-dependent density functional theory* in its explicit *real-time* propagation form (RT-TDDFT), it is now possible to investigate the quantum dynamic response of electrons in systems of great chemical complexity, such as DNA in water, as required here for studying electronic stopping. Using large-scale RT-TDDFT simulations [21-23], we show here that high energy protons transfer significantly more energy to the sugar phosphate side chains than the nucleobases of DNA, generating highly energetic holes on the side chains as key source of oxidative damage.

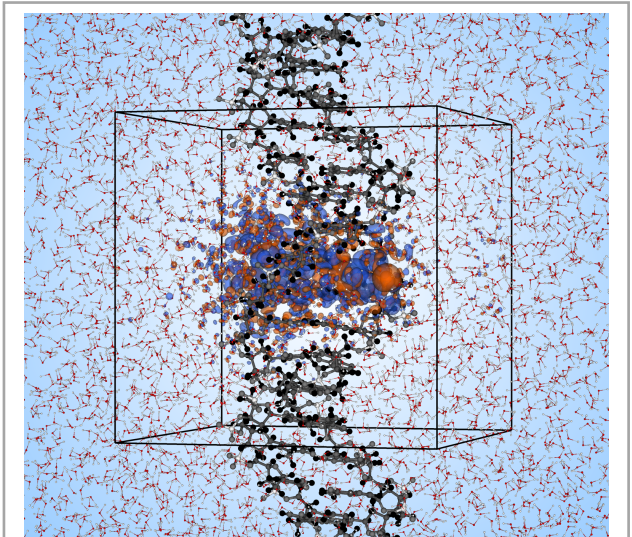

FIG. 1. Simulation cell for solvated DNA. The simulation cell, outlined in black, is shown with periodic boundary conditions for solvated DNA. Blue (orange) isosurfaces represent decreases (increases) in electron density in response to a proton moving through the center of DNA at 0.50 a.u. velocity (6.25 KeV).

**Results and Discussion**

Figure 1 shows B-DNA (i.e. normal right-handed DNA [24]), solvated in water, with the simulation cell outlined by the black box. The DNA strand within the simulation cell comprises one full turn of the double helix. Including the surrounding water molecules, the dynamics of more than 11,500 electrons are explicitly simulated as they respond to an irradiating proton. Additional computational details are discussed in the Computational Method section of the Supplemental Material [25]. In our previous work on dry DNA [26], this first-principles approach was used and compared to the widely-used semi-empirical perturbation theory formalism, based on the dielectric function [13], showing good agreement. We consider two paths for an irradiating proton as shown in Fig. 2A; the Base path directly through the center of the DNA molecule (shown in cyan in Fig. 2A and Fig. S1), and the Side path along the sugar-phosphate side chain (shown in red in Fig. 2A and Fig. S1). Simulations were performed at six different proton kinetic energies (0.5 to 6.0 a.u. velocity, or equivalently 6.25 to 900 keV kinetic energy) for each path, including velocities close to the Bragg peak in dry DNA [26] and liquid water [1]. All atoms, other than the irradiating proton, are fixed in place to study the electronic stopping phenomenon here, and the time scale of each simulation (0.27 to 3.38 fs, depending on proton velocity) is too short for any significant nuclear motion [27]. The energy transfer rate, referred to as electronic stopping power, can be obtained as a function of the proton velocity for each individual path [28]. It is convenient to express the stopping power in terms of the work done on the non-equilibrium system of electrons by a single "projectile" proton, and the total electronic energy change of the system can be used in practical computation of the stopping power (see Supplemental Material for details [25]) [21]. Comparison of the solvated DNA stopping power curves (Fig. 2B, solid lines) reveals that the stopping power magnitude for the Side path is more than three times larger than that for the Base path at the peak, and at least twice as large at all velocities. This difference increasingly diminishes with higher proton velocities. While the Bragg peak positions for both paths remains similar to that of liquid water [1] (Fig. 2B, black line), the stopping power magnitude for the Side path is 40% larger at the Bragg peak. This is of particular importance as the electronic stopping power for liquid water is generally used for calibrating proton beam in radiation oncology [29]. We also note that at the higher velocities of 4.00 and 6.00 a.u., the stopping power magnitude for the Side path is nearly identical to that of liquid water. Compared to the case of dry DNA [26], the Side path also shows much larger electronic stopping power for solvated DNA (see Fig. S3). Negative charges, specifically lone-pair electrons on phosphate groups on the DNA side chains, were found to be largely responsible for this difference (see Supplemental Material for details [25]). In order to gain molecular-level insights in this complex system, we employ the time-dependent maximally-localized Wannier function (TD-MLWF) gauge [30, 31]. TD-MLWFs are spatially localized on different chemical moieties, creating a chemically intuitive picture of the DNA-water electronic system. Geometric centers of the TD-MLWFs, commonly referred to as Wannier centers (WCs), are shown in Fig. 2A. The TD-MLWFs can be grouped into different chemical subgroups, and the electronic response of DNA can be separated from that of the solvating water molecules. The response is further studied in terms of DNA chemical moieties, nucleobases and sugar phosphate side chains, by analyzing changes to the spatial spread (Wannier center variance) and Wannier center displacement of individual TD-MLWFs. Figure 3 shows the Wannier center displacements (A) and spread changes (B) for the two paths at the proton velocity of 1.64 a.u. (67.19 keV), close to the Bragg peak. For both paths, greater than 80% of the Wannier center displacements are within 10 a.u. of the proton path, and more than 90% of the spread changes are within 5 a.u. of the proton path; the electronic excitation response is highly localized near the path of the irradiating proton. In Fig. 3, the hatched areas indicate contributions from the sugar phosphate side chain. The response for the Side path is almost entirely from the phosphate side chain, greater than 87% of the displacements and more than 98% of the spread changes, while the Base path shows minimal contribution from the phosphate side chain, with over 75% of the displacements and more than 90% of the spread change from nucleobases. These key excitation features are also observed at higher and lower velocities (see Figs. S6, S7 in Supplemental Material [25]). Our simulations show that the sugar phosphate side chain molecules absorb much more energy than nucleobases in the proton beam.

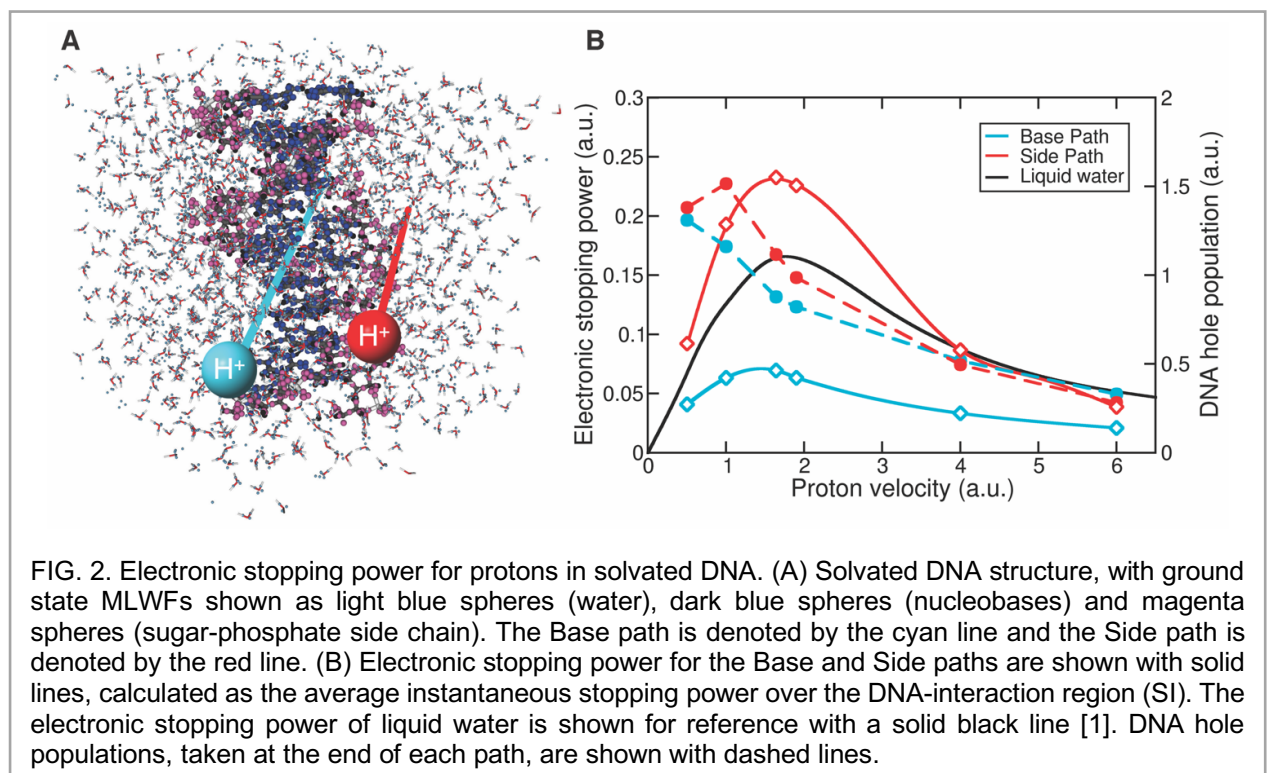

FIG. 2. Electronic stopping power for protons in solvated DNA. (A) Solvated DNA structure, with ground state MLWFs shown as light blue spheres (water), dark blue spheres (nucleobases) and magenta spheres (sugar-phosphate side chain). The Base path is denoted by the cyan line and the Side path is denoted by the red line. (B) Electronic stopping power for the Base and Side paths are shown with solid lines, calculated as the average instantaneous stopping power over the DNA-interaction region (SI). The electronic stopping power of liquid water is shown for reference with a solid black line [1]. DNA hole populations, taken at the end of each path, are shown with dashed lines.

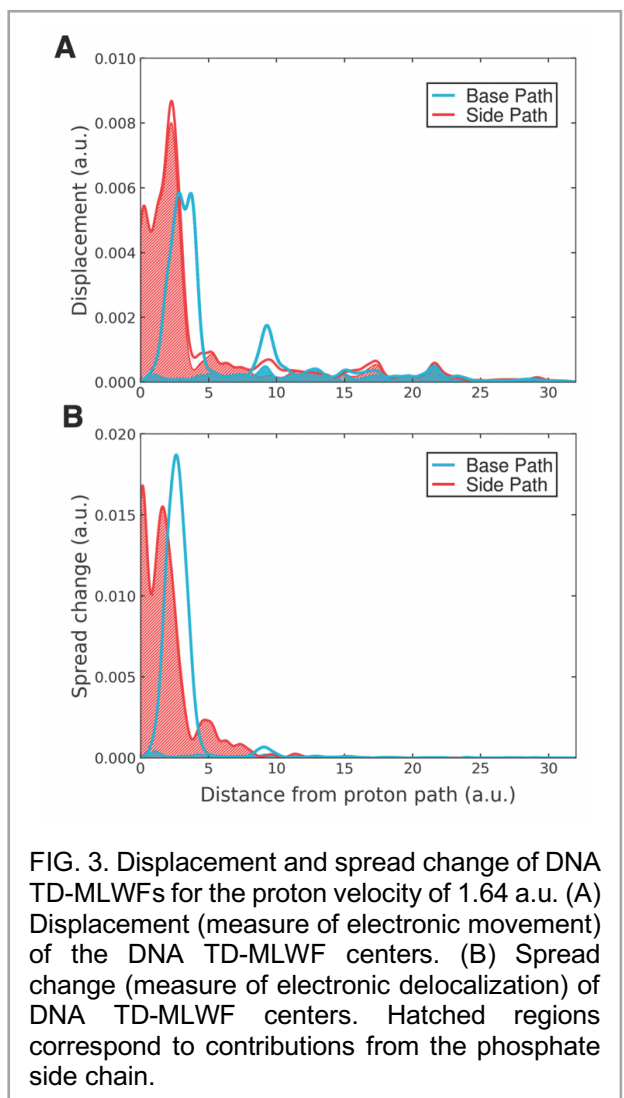

FIG. 3. Displacement and spread change of DNA TD-MLWFs for the proton velocity of 1.64 a.u. (A) Displacement (measure of electronic movement) of the DNA TD-MLWF centers. (B) Spread change (measure of electronic delocalization) of DNA TD-MLWF centers. Hatched regions correspond to contributions from the phosphate side chain.

Electronic stopping power is often thought to be directly proportional to electronic excitations, or more specifically the number density (i.e. population) of holes (or excited electrons) generated under ionizing radiation [32]. Figure 2B also shows the formation of holes on DNA as a function of the irradiating proton velocity (dashed lines). The DNA hole populations were found to reach a constant value by the end of each simulation trajectory, and charge transfer from DNA to the irradiating proton does not contribute to the hole population (see Supplemental Material for details [25]). While the stopping power is considerably different between the two paths, the hole population is only slightly larger for the Side path. For the 1.00 a.u. proton velocity, where the largest difference in DNA hole population is observed, 1.3 times as many holes are generated for the Side Path relative to the Base Path. However, the stopping power is more than three times greater for the Side Path at the same velocity, relative to the Base Path. Therefore, the differences in electronic stopping power cannot be explained simply by the number density of holes formed on DNA. The stopping power also depends on the energetics of the generated holes. To quantify the energetics, we project the DNA-localized TD-MLWFs onto the energy eigenstates. Figure 4 shows the hole population on DNA as a function of energy for the Base path (A) and the Side path (B). The electronic density of states (DOS) is also shown as a reference (dashed line). A significant number of holes are formed in the deeper lying states for the Side path, and essentially no holes are formed within approximately 2 eV of the highest occupied molecular orbital (HOMO), which is aligned at 0 eV in Fig. 4. The HOMO in DNA largely comprises nucleobase electronic states, and the Base path shows a sharp peak close to HOMO, which is responsible for 10-15% of the holes generated on DNA, depending on the irradiating proton velocity. The deeper-lying DNA-states, at around -20 eV, largely derive from the DNA sugar-phosphate side chains. For the Side path, as much as 8% of the total DNA holes are generated between -20 and -25 eV. At the same time, holes generated in this energy range represent only 2% or less of the total hole population for the Base path, depending on the proton velocity. This characteristic difference in hole energetics is largely responsible for the significant difference in the stopping power for these two paths, and more extensive DNA phosphate side chain damage can be expected than DNA nucleobase damage, under proton irradiation. Additionally, at velocities away from the Bragg peak (e.g. 4.00 and 0.50 a.u., above and below the Bragg peak respectively), the hole generation in the deeper-lying regions (corresponding to the sugar-phosphate side chains) becomes quite small, as seen in Fig. 4. Thus, significant strand damage to DNA can be expected only for proton velocities close to the Bragg peak.

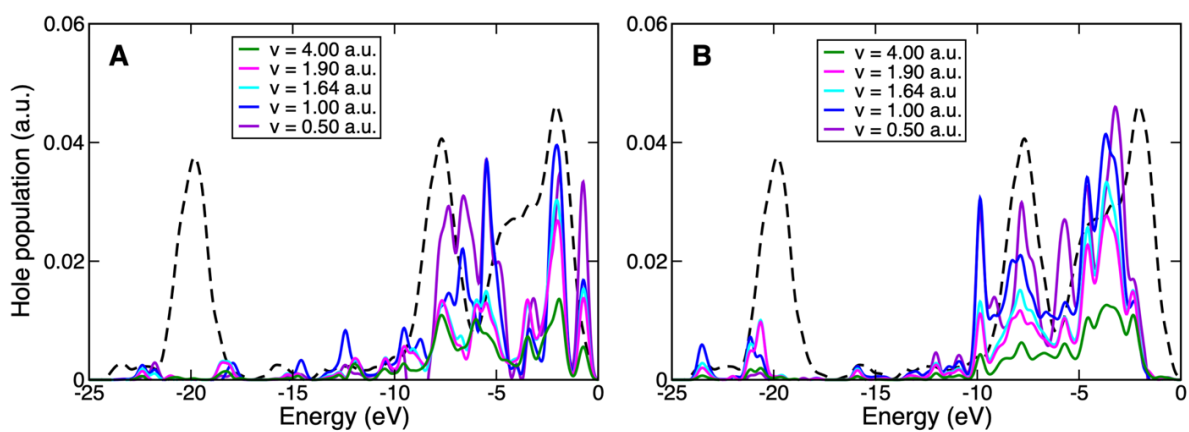

FIG. 4. DNA hole populations as a function of energy for the Base (A) and Side (B) paths. DNA TD-MLWFs are projected onto the eigenstates of the system at equilibrium to calculate the energies at which holes are generated in DNA at the end of simulations. For reference, the density of states is shown with a dashed line. Gaussian broadening of 0.25 eV was used for all hole energy distributions. HOMO is aligned to be at 0 eV. Nearly identical energetics were observed at the end of the DNA-interaction region (see Figs. S10, S11 in Supplemental Material).

**Conclusion**

The electronic excitation response of DNA to high-energy protons in water was investigated. Quantum-mechanical simulations revealed intricate molecular-level details of the energy transfer process from the high-energy protons to DNA in water. With proton irradiation, significantly more energy was deposited onto the sugar-phosphate side chains rather than onto the nucleobases. The enhanced energy transfer to the DNA strands derives from the generation of highly energetic holes on the side chains. These highly energetic holes are a key source of oxidative damage, and their formation on the side chains is likely the source of DNA strand damage. The first-principles simulation results presented here fill a key knowledge void in understanding detailed mechanisms for extensive DNA strand break lesions observed with a proton beam. In the context of proton beam cancer therapy, the present work will add to the growing knowledge base for building increasingly more sophisticated multi-scale modeling in medical physics [16, 33, 34].

**Acknowledgments/Funding:**
The authors would like to thank Dr. Yi Yao for helpful discussions. The work is supported by National Science Foundation grant CHE-1565714 (D.C.Y. and Y.K.), CHE-1954894 (C.S. and Y.K.), and DGE-1144081 (D.C.Y.). The Qb@ll code used in this work implements theoretical formalisms developed under NSF grant no. OAC-2209858 and CHE-1954894. An award of computer time was provided by the Innovative and Novel Computational Impact on Theory and Experiment (INCITE) program. This research used resources of the Argonne Leadership Computing Facility, which is a DOE Office of Science User Facility supported under contract DE-AC02-06CH11357.

## Supplemental Material for:

Electronic Excitation Response of DNA to High-energy Proton Radiation in Water
Christopher Shepard[1], Dillon C. Yost[3], Yosuke Kanai[1,2*]

1. Department of Chemistry, University of North Carolina at Chapel Hill; Chapel Hill, NC, 27514, USA.
2. Department of Physics and Astronomy, University of North Carolina at Chapel Hill; Chapel Hill, NC, 27514, USA
3. Department of Materials Science and Engineering, Massachusetts Institute of Technology; Cambridge, MA, 02139, USA.

*Corresponding author. **Email:** ykanai@unc.edu

# 1. Computational Method

**Molecular dynamics generation of solvated DNA structure**
The DNA structure was first generated from a strand of B-DNA by performing classical molecular dynamics simulation using the Amber MD code [1, 2]. The DNA strand contains the same 10 base pairs (sequence CGCGCTTAAG) as in our previous work [3], comprising one full turn of the double helix. This ensures the periodic images in the z-direction were commensurate with the periodicity of the macromolecule. A cubic simulation cell, with dimensions of 34.43 angstrom (A), was used as it was large enough for the DNA strands in the z-direction, while preventing errors from the periodic images in the x and y directions [3]. The DNA strand was subsequently solvated with 1119 explicit water molecules, using GROMACS [4]. The Parmbsc1 force field [5] was used for DNA, and the SPC/E [6] force field was used for water molecules. Energy minimization was then performed using the steepest descent minimization until the force went below 1,000.0 kJ/mol/nm. The system was subsequently equilibrated for 10 ns using an NVT ensemble, with all DNA atoms restrained to their initial positions using the LINCS algorithm [7]. A time step of 2 fs was used to propagate the equations of motion and the Particle Mesh Ewald (PME) method [8] was used for long-range electrostatic interactions. Initial velocities were assigned using a Boltzmann distribution at 300 K and temperature coupling was achieved using the modified Berendsen thermostat [9], reference temperature of 300 K and time constant of 5 ps. After equilibrating the structure for a few nanoseconds in the MD simulation, a snapshot was taken for RT-TDDFT simulations.

**Details of RT-TDDFT simulations**
For all simulations described in the main text we use the Qb@ll [10] branch of the Qbox code [11], with our implementation of real-time time-dependent density functional theory (RT-TDDFT) based on the plane-wave pseudopotential formalism [12]. The simulation includes 3991 atoms (634 for DNA, 3357 for water) and 11,172 electrons, necessitating the use of the highly scalable and massively parallel Qb@ll code [13] to perform this ab initio study. Additionally, all simulations required up to 262,144 Intel KNL 7230 cores on the Theta supercomputer at the Argonne leadership Computing Facility. A study of this scale was only possible with such peta-scale resources. All atoms, including the irradiating proton, were represented by norm-conserving Hamann-Schluter-Chiang-Vanderbilt (HSCV) pseudopotentials [14, 15]. In all RT-TDDFT simulations the time-dependent Kohn-Sham orbitals are expressed in the maximally localized Wannier function (MLWF) [16] gauge and propagated as time-dependent (TD)-MLWFs [17, 18] in order to study the induced excitations on the molecular-level. An 2.0 attosecond time step was used in the enforced time reversal symmetry (ETRS) [19] propagator for time integration of the electronic states. A planewave cutoff energy of 50 Rydberg energy was used, and the Perdew-Burke-Ernzerhof (PBE) functional [20] was used for the exchange-correlation (XC) approximation. PBE, a generalized gradient approximation (GGA) functional, is known to give artificial charge delocalization in some cases [21]. However, in the context of electronic stopping simulations, earlier RT-TDDFT studies compared the PBE functional to different hybrid functionals, such as PBE0 [22], and found essentially identical results for calculation of various properties, including ion effective charge and stopping power [23, 24]. In its initial state in the RT-TDDFT simulation, the projectile/irradiating proton is fully ionized (i.e. $H^+$). At each step the proton position is updated based on a specified velocity, which is held constant throughout the simulation for the purpose of calculating electronic stopping power. The positions of all other atoms are held constant as the electron density changes in response to the irradiating proton until the proton reaches the end of its trajectory and the simulation is stopped.

## 2. High-energy Proton Trajectories

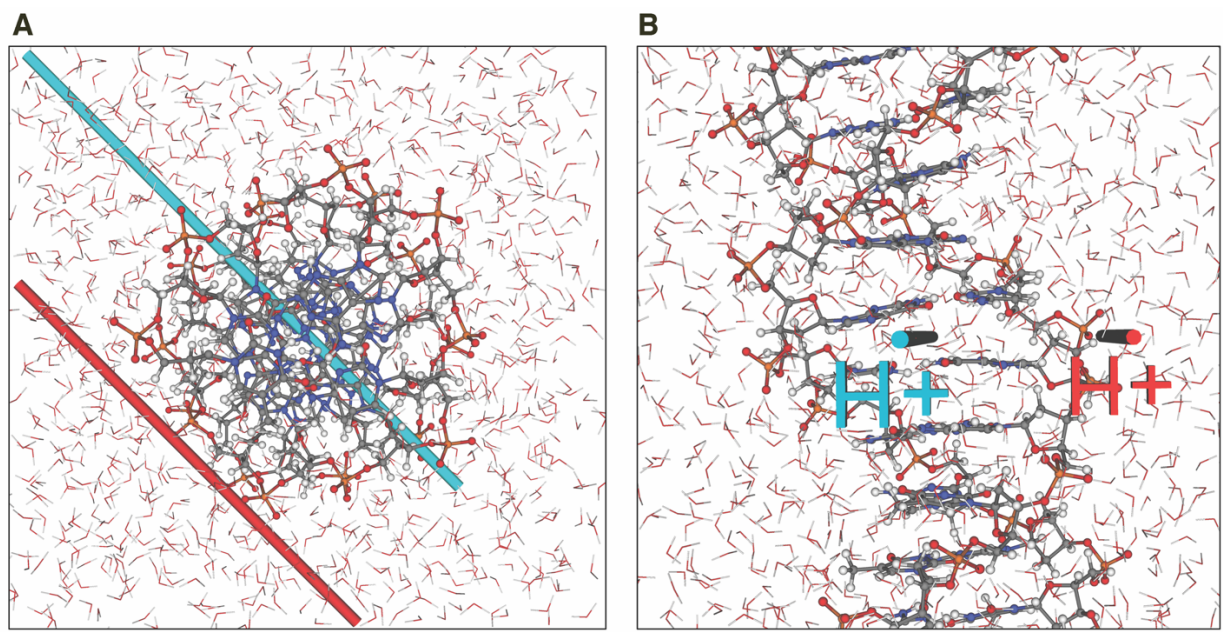

**Fig. S1.**
**Enhanced view of high-energy proton trajectories.** The above schematic shows the two proton paths studied in this work. **(A)** Top-down view of solvated DNA, with the cyan line representing the Base path and the red line representing the Side path. **(B)** Side view of the two trajectories, with the Base path (cyan) through the center of DNA and the Side path (red) just outside the DNA sugar-phosphate side chain.

# 3. Calculation of Electronic Stopping Power

In the RT-TDDFT simulations, the projectile proton is moved at a constant velocity in a specified direction. Thus, in this non-equilibrium simulation, additional work is done by the projectile proton on the system, and the total energy of the solvated DNA system is not conserved. This increase in total energy, from the work done by the projectile proton, can be then used as a quantitative metric to measure the electronic excitation response [12, 25, 26]. In particular, the rate of energy transfer from the projectile to electrons in a target matter is measured per unit distance of the traveling projectile as a function of the projectile velocity. This quantity is called electronic stopping power in science or linear energy transfer in medicine. For the purpose of determining the electronic stopping power for only the DNA portion within solvated DNA, dry DNA (DNA in vacuum) is also simulated using the same projectile proton paths. For dry DNA, all electronic energy change can be attributed to DNA, as no energy change occurs within the vacuum region. Thus, a DNA-interaction region can be defined as where the projectile proton is interacting with the DNA strand. We define the DNA-interaction region by local maxima in the instantaneous stopping power curve, calculated as electronic energy change with respect to the proton displacement, for dry DNA, as seen in Fig. S2. The average of instantaneous stopping power over this region gives the stopping power for DNA in water.

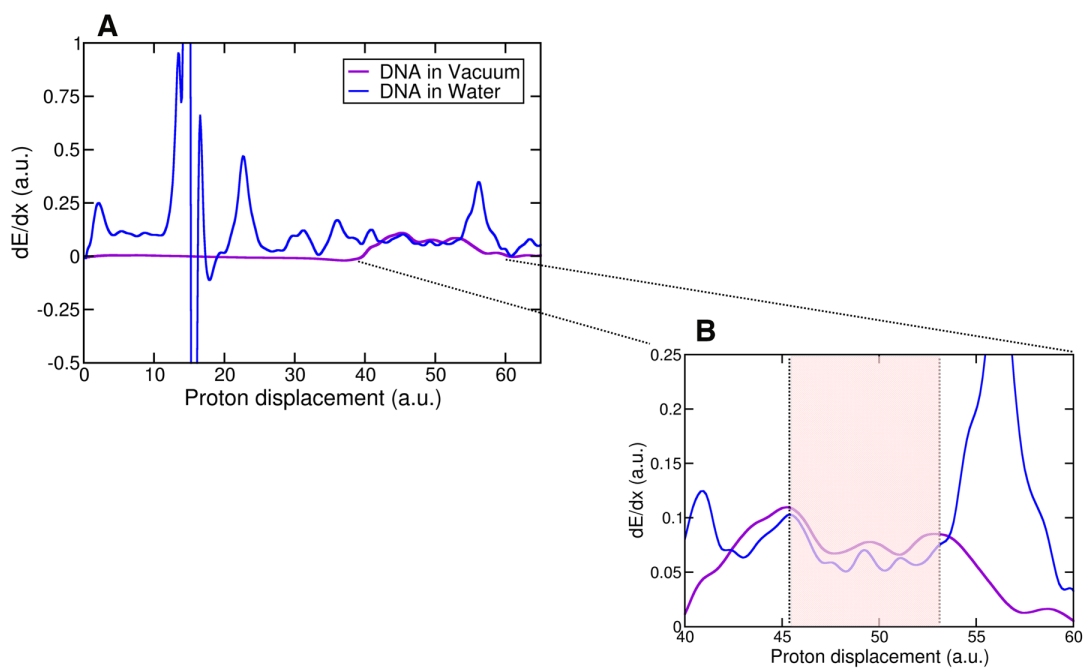

**Fig. S2.**
**Determination of electronic stopping power region for DNA. (A)** Instantaneous stopping power (dE/dx) as a function of the proton displacement for a proton moving at 1.64 a.u. (stopping power maximum for dry DNA) velocity along the Base path. **(B)** Zoomed-in view of the region where the projectile proton is close to DNA, with the hatched portion showing the DNA-interaction region for the 1.64 a.u. velocity along the Base path.

# 4. Effect of Deprotonation of Phosphate Groups in DNA on Electronic Stopping Power

The extent of the deprotonation for DNA in water simulated here corresponds to DNA at the physiological pH of approximately 7.4 [27]. Under the physiological condition, the phosphate groups have a p$K_a$ close to 2, and therefore carry a pair of unbonded electrons [28]. The comparison of the electronic stopping power of DNA in water to that of dry DNA [3] is shown in Fig. S3. To understand the effect of having unbonded lone-pair electrons (i.e. due to deprotonation), we also compare the electronic stopping power of dry DNA with fully protonated phosphate groups [3] to dry DNA with deprotonated phosphate groups (structures shown in Fig. S4). As discussed in the main text, the Base path showed minimal difference in stopping power, while the Side path showed a significant increase in stopping power for the deprotonated structure, indicating the unbounded lone-pair electrons on the sugar-phosphate side chains are largely responsible for the difference, as seen in Fig. S5.

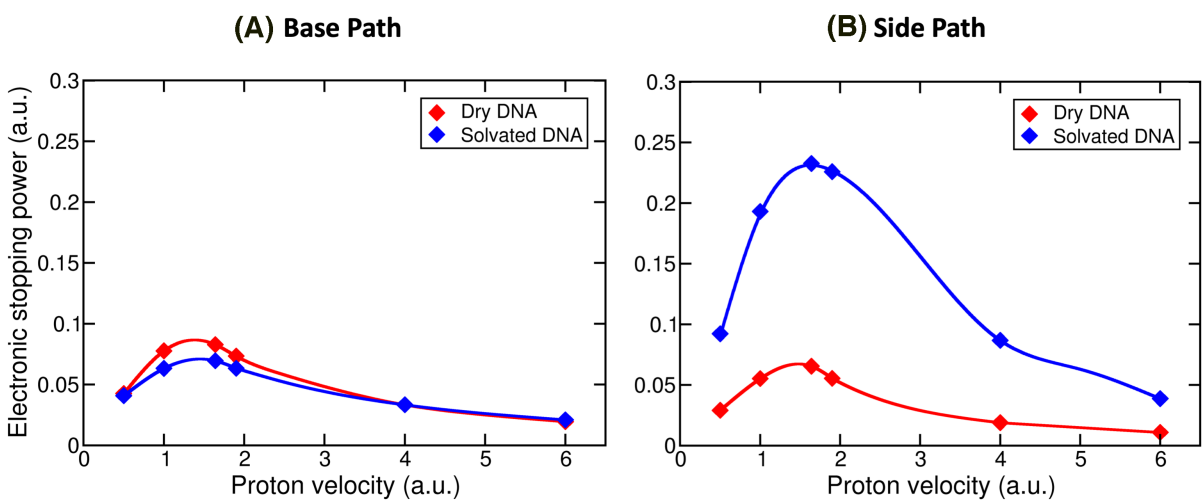

**Fig. S3.**

**Electronic stopping power comparison to dry DNA.** Comparison of the electronic stopping power for solvated DNA to that of dry DNA for the **(A)** Base path and **(B)** Side path. The dry DNA structure used here, for comparison, has fully protonated sugar-phosphate side chains [3].

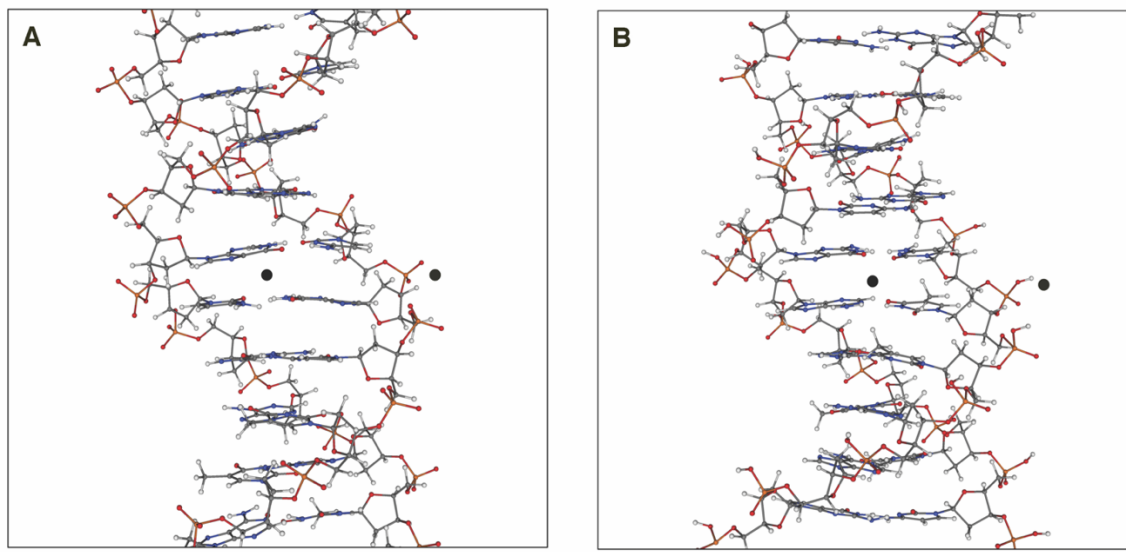

**Fig. S4. Comparison of dry DNA structures.** (A) Deprotonated and (B) protonated dry DNA structures used for electronic stopping power comparison (shown in Fig. S5), with paths (denoted by black dots).

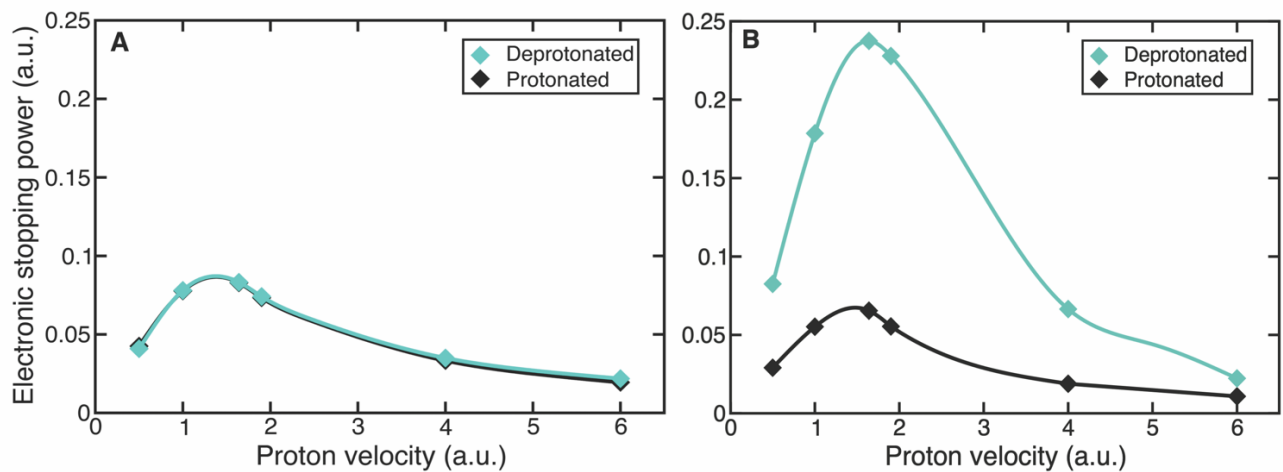

**Fig. S5.**

**Electronic stopping power for dry DNA. (A)** Electronic stopping power for the Base path, and **(B)** electronic stopping for the Side path, for "dry" DNA in vacuum with protonated and deprotonated phosphate groups on the side chains.

## 5. DNA Wannier Center Displacements and Spread Changes

For proton velocities higher (Fig. S6) and lower (Fig. S7) than the Bragg peak, highly localized electronic excitation is observed along the projectile path, as seen in Fig. 3 of the main text. The same key excitation features as seen in Fig. 3 are also observed here, as the Side path excitation is mostly from the phosphate side chain, while the Base path excitation shows only small contributions from the phosphate side chain.

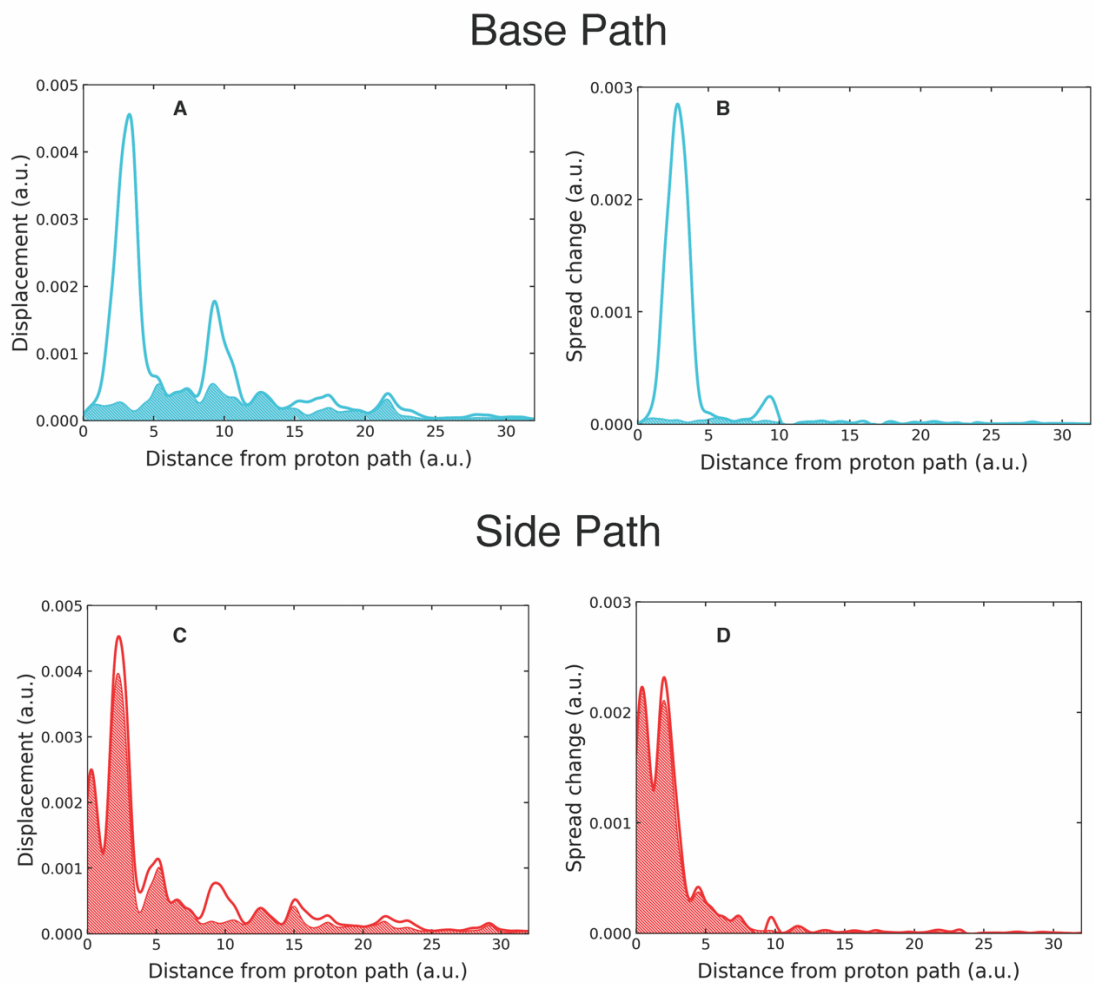

**Fig. S6.**

**Displacement and spread change of DNA TD-MLWFs for the proton velocity of 4.00 a.u.**
**(A)** Displacement of DNA Wannier centers (i.e. geometric centers of TD-MLWFs) for the Base path. **(B)** Spread change of DNA Wannier centers for the Base path. **(C)** Displacement of DNA Wannier centers for the Side path. **(D)** Spread change of DNA Wannier centers for the Side path. Hatched regions correspond to contributions from the phosphate side chains.

## Base Path

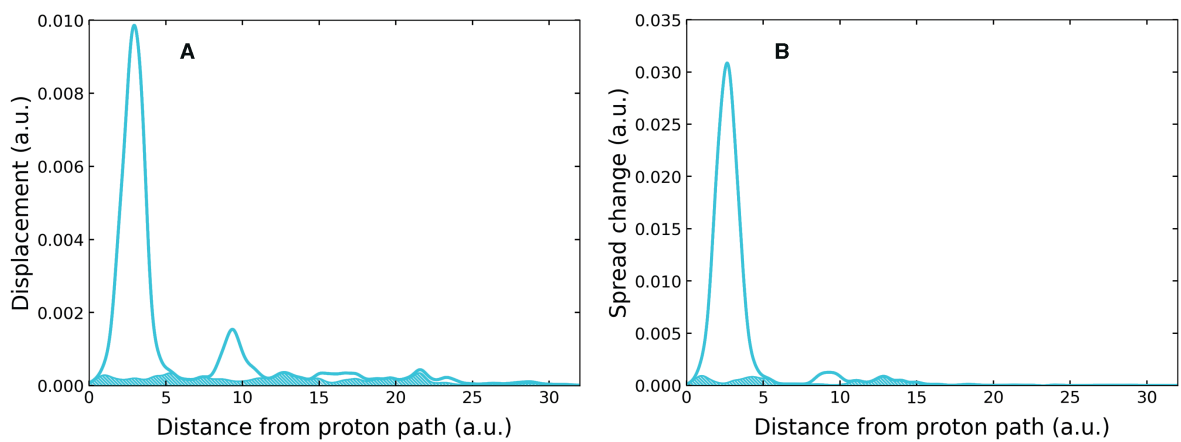

## Side Path

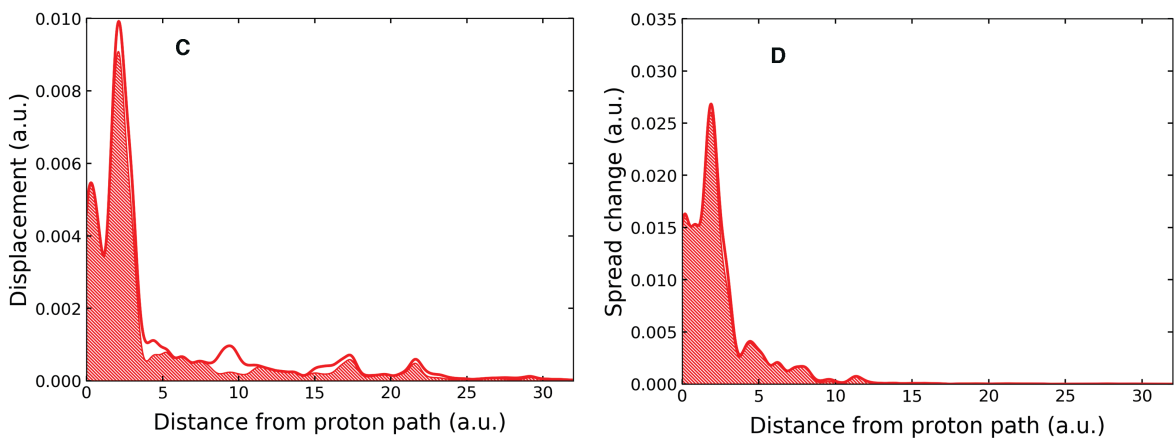

**Fig. S7.**

**Displacement and spread change of DNA TD-MLWFs for the proton velocity of 1.00 a.u. (A)** Displacement of DNA Wannier centers for the Base path. **(B)** Spread change of DNA Wannier centers for the Base path. **(C)** Displacement of DNA Wannier centers for the Side path. **(D)** Spread change of DNA Wannier centers for the Side path. Hatched regions correspond to contributions from the phosphate side chains.

# 6. Time Dependence of DNA Hole Populations

Time-dependent DNA hole populations as a function of proton displacement for the Base path (top panels) and Side path (bottom panels). For both paths, the total DNA hole population (purple) remains negligibly small until the projectile comes within close proximity of DNA, where it begins to quickly increase. The DNA hole population levels off and remains approximately constant once the projectile proton leaves the vicinity of DNA. Additionally, decomposition of the DNA hole populations in terms of holes formed on the nucleobase pairs (red) vs. the phosphate side chain (orange) shows the same key excitation features observed in Fig. 3 and Supplementary Figures S6 and S7.

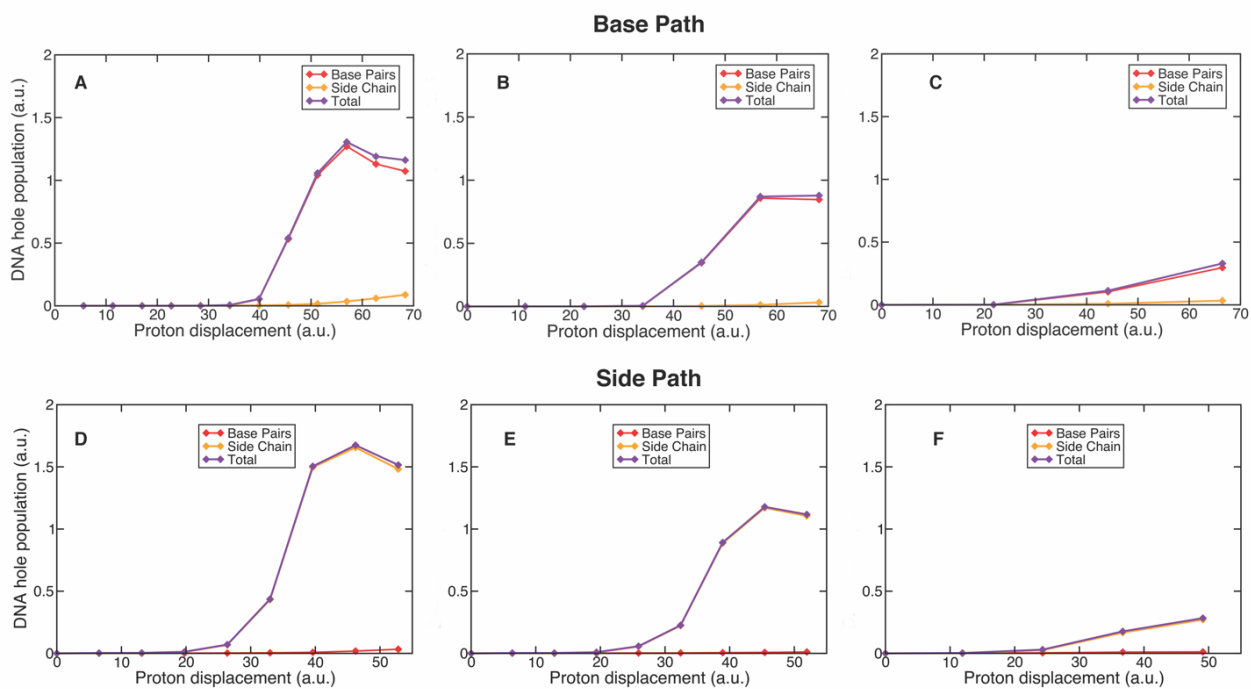

**Fig. S8.**

**Time dependence of DNA hole populations. (A, B, C)** DNA hole populations for the Base path with the projectile proton at velocities of 1.00, 1.64 and 6.00 a.u., respectively. **(D, E, F)** DNA hole populations for the Side path with the projectile proton at velocities of 1.00, 1.64 and 6.00 a.u., respectively.

## 7. Charge Transfer to Projectile Proton

Voronoi charge partitioning [29] of the electron density was used to determine electronic charge on the projectile proton throughout the simulations. At high velocities of 4.00 a.u. and 6.00 a.u., the proton remains essentially as a bare ion. At lower velocities, the projectile proton gains sizable amounts of charge, a trend consistent with previous studies of liquid water [23] and dry DNA [3], as seen in Fig. S9. However, unlike for dry DNA, the projectile proton was found to reach an effective charge state prior to interaction with the DNA strand, as it moves through water. Thus, the charge transfer (CT) contribution to the DNA hole population was considered negligible for both paths.

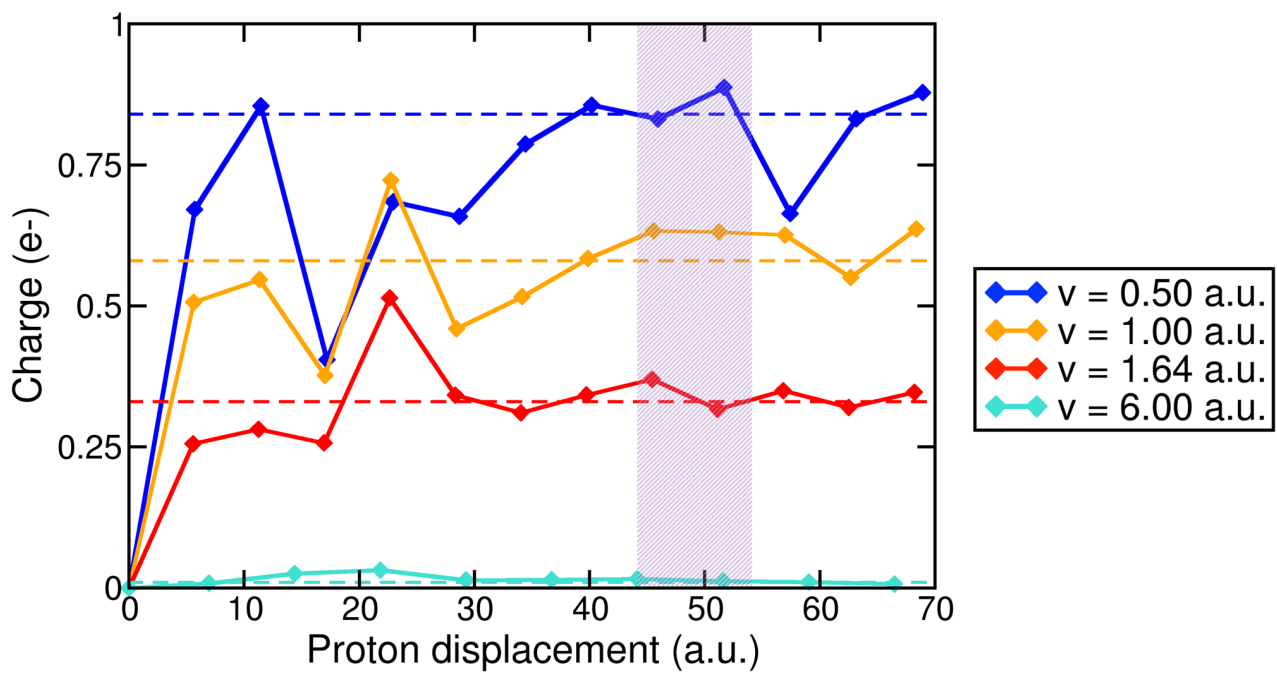

**Fig. S9.**

**Charge transfer to the projectile proton.** Effective charge accumulated on the projectile proton, as a function of the proton displacement, for a representative set of velocities along the Base path. The DNA-interaction region is shown hatched in violet. Dashed lines represent the charge captured by projectile protons in bulk water [30]. Identical trends are observed for the Side path.

## 8. Time Dependence of DNA Hole Population Energetics

DNA hole population energetics for the proton at the end of the DNA-interaction region (orange) and for the proton at the end of the path (green) for the Base path (Fig. S10) and the Side path (Fig. S11). As seen below, nearly identical energetics are observed, with only slight differences in magnitude for the low (0.50 and 1.00 a.u.) velocities.

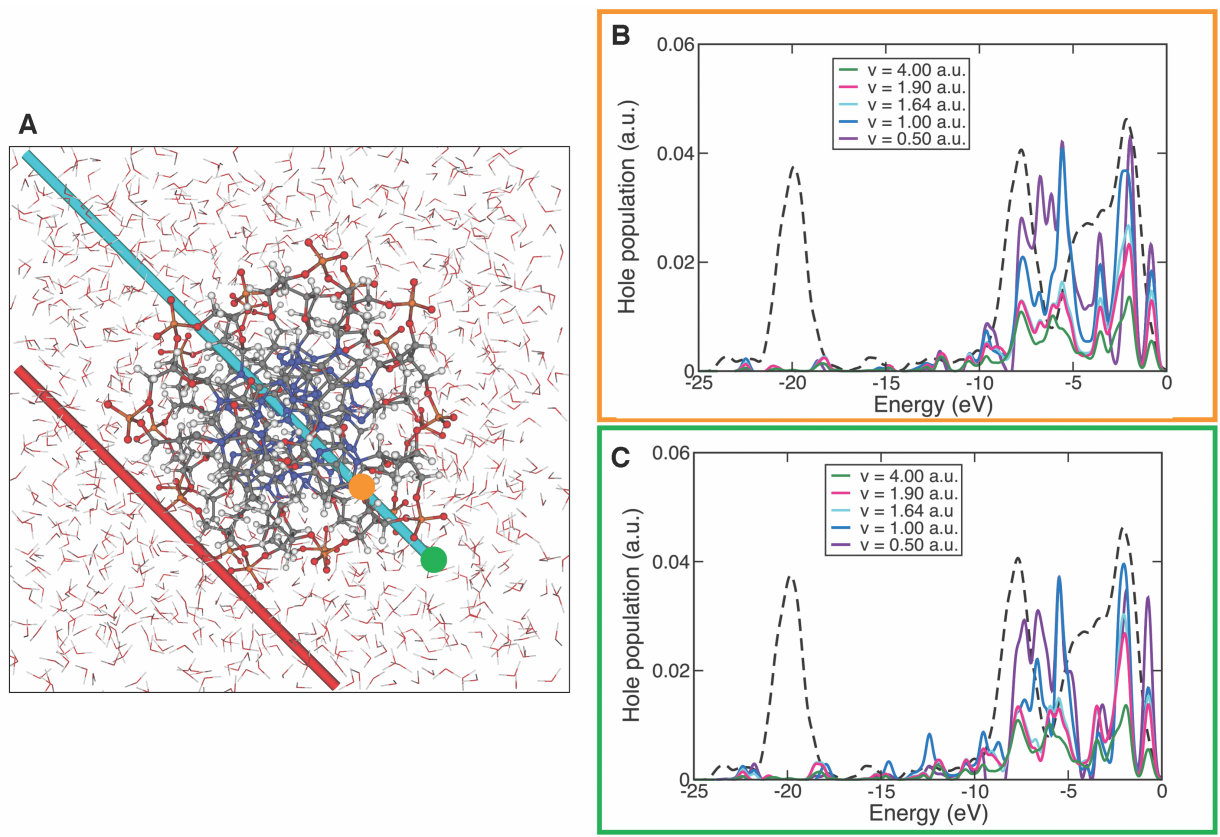

**Fig. S10.**

**Time dependence of DNA hole population energetics for the Base path. (A)** Points of comparison (orange: end of DNA-interaction region, green: end of path) for projections of DNA TD-MLWFs onto the energy eigenstates along the Base path. **(B)** DNA hole energetics at the end of the DNA-interaction region. **(C)** DNA hole energetics at the end of the path.

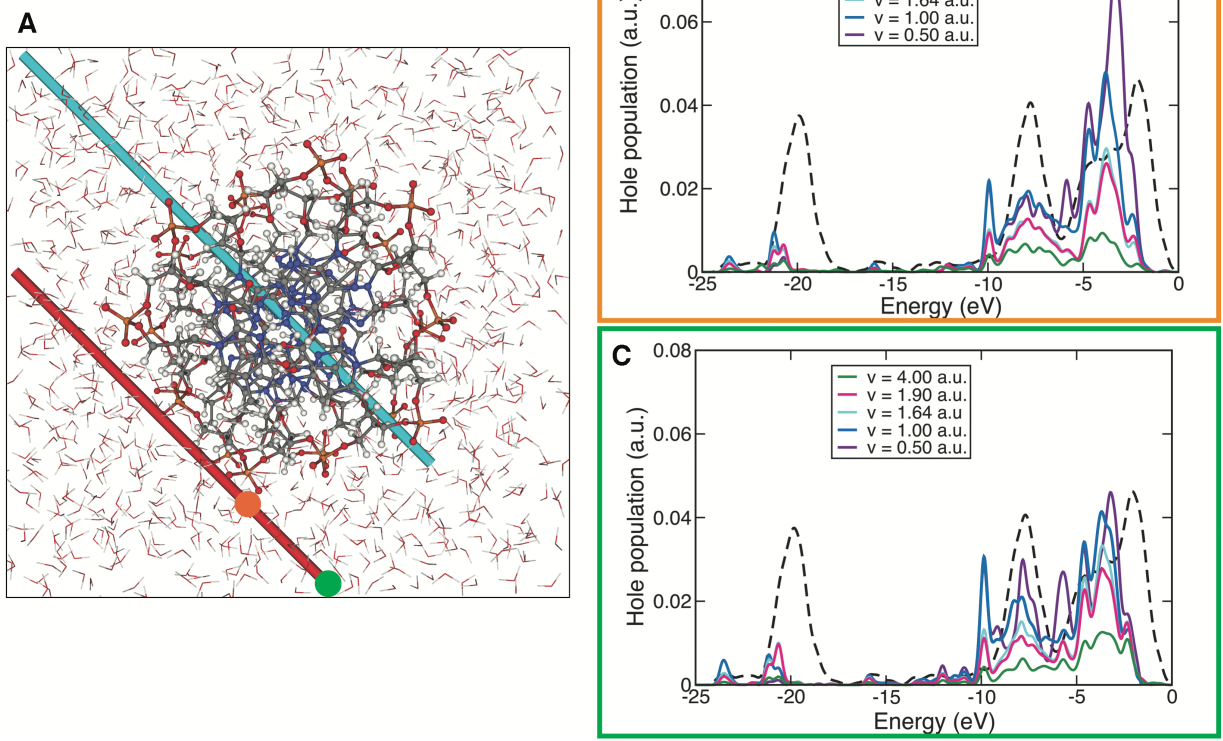

**Fig. S11.**

**Time dependence of DNA hole population energetics for the Side path. (A)** Points of comparison (orange: end of DNA-interaction region, green: end of path) for projections of DNA TD-MLWFs onto the energy eigenstates along the Side path. **(B)** DNA hole energetics at the end of the DNA-interaction region. **(C)** DNA hole energetics at the end of the path.

# SI References